 \documentclass[preprint,12pt]{elsarticle}

\usepackage{amssymb}
\usepackage{comment}

\usepackage{lineno}

\journal{Nuclear Physics A}

\begin{document}

\begin{frontmatter}



\title{Relativistic four-nucleon calculations with rank-one separable potential}

\author[label1]{Serge  Bondarenko}
\author[label1]{Sergey Yurev}\ead{yurev@jinr.ru}
\address[label1]{BLTP, Joint Institute for Nuclear Research, Dubna, 141980, Russia }

\begin{abstract}
A solution to the relativistic generalization of the four-particle integral Faddeev-Yakubovsky equation  
is carried out.
Only states with zero orbital angular momentum, $S$ states, are considered in the calculations.
A rank-one separable potential for nucleon-nucleon interaction is used to solve the two-nucleon 
Bethe-Salpeter equation.
Calculations are carried out taking into account both "3+1" and "2+2" subchannels in the equation.
The system of integral equations is solved by the iteration method and the binding energy
of the helium-4 nucleus is found.
The calculated results are compared with experimental data and nonrelativistic.
\end{abstract}




\end{frontmatter}



\section{Introduction}
\label{intro}

The increasing of the energies in particle and nuclear accelerators poses the question of relativistic methods for studying nuclear systems. 
In particular, this also applies to problems in which few-body systems are studied.
The need to study nuclei with a few nucleons is due, among other things, to the fact that they are the simplest systems in which it is possible to take into account the influence of three- and four-particle forces.
These systems are also important for studying the evolution of nucleon-nucleon interaction 
with an increase in the number of particles in the nucleus.
The well-known methods such as the Faddeev~\cite{Faddeev} and Yakubovsky~\cite{Yakubovsky:1966ue} equations can not be applicable due to their obvious non-relativistic nature.

There are a number of attempts to generalize non-relativistic equations to
the relativistic case. These include the approaches of Gross~\cite{Gross} and Glockle group~\cite{W}  for systems of
three particles and Kamada~\cite{Kamada:2019dcn} for four ones. These approaches, despite their
relativistic nature, have limitations that do not allow a full consideration of
relativistic effects. 
For example, in Gross's approach there is only one particle off mass shell.
In contrast this, in~\cite{Rupp:1987cw}, when constructing the Bethe-Salpeter-Faddeev equation, it is assumed that
all three particles of the system are off mass shell.
In approach~\cite{W}   despite the use of the Poincare principle 
and the formulas of the special theory of relativity
the dynamic variables is the usual 3-momenta.
Same like the potential~\cite{H}  and the Green's function and the two-particle t matrix dependence.
In this paper, the ideas~\cite{Rupp:1987cw} of constructing an approach
to studying few-body systems based on the Bethe-Salpeter (BS) equation~\cite{BS}
are developed.
Despite the conclusion of the article~\cite{W}  that in their approach relativistic effects 
are comparable to the influence of three-particle forces
we believe that considering equations with full 4-momenta dynamics 
is more important than taking into account three-particle forces.
 Although in the future the approach we use will be generalized 
to the case of taking into account three-particle forces.
And at this stage of research we use only the pair potential
as it was~\cite{Faddeev,Yakubovsky:1966ue} in the original non-relativistic approaches.

The relativistic Bethe-Salpeter-Faddeev equations were successfully applied to describe three-nucleon systems~\cite{Bondarenko:2020baw,Bondarenko:2021lmq}. It allows one to extend the formalism to a relativistic
generalization of the Faddeev-Yakubovsky (FY) equation for studying relativistic four-nucleon systems.

In the previous work~\cite{Bondarenko:2024mkl} we constructed this generalization and obtained its solution using the rank-one separable potential of nucleon-nucleon ($NN$) interaction in the Yamaguchi form.
In doing so, we made a number of assumptions to simplify the calculations.
In particular, we did not take into account the "2+2" subchannel. 

In this work, we continue the research 
of the previous work on the topic under consideration.
As the $NN$-interaction potential, we use the Yamaguchi rank-one separable potential in monopole and dipole forms.  In numerical calculations, we limit ourselves to considering only the paired $S$-state.
Unlike our previous work, in this article
we solve the full equation taking into account the
"3+1" and "2+2" subamplitudes. 

The paper is organized as follows: Sec. 2 gives the potential of nucleon-nucleon interaction, 
Sec. 3 defines the relativistic generalization of the Faddeev-Yakubovsky equation, 
Sec. 4 discusses the calculations and results,
and finally Sec. 5 gives the conclusion.

\section{Rank-one separable relativistic potential of $NN$-interaction}

In this paper, a four-nucleon system is considered taking into account only paired $NN$-interactions, thus three- and four-nucleon forces are not taken into account while solving the FY equation. Also, nucleons are described by relativistic scalar propagators.

The nucleon-nucleon interaction potential, which is used to solve the BS equation and then the Bethe-Salpeter-Faddeev equation, is taken in the separable form:
\begin{eqnarray}
v(k , k' ) = \lambda 		{g}(k) 		{ g}(k'),
\label{separ_NN}
\end{eqnarray}
where the function $g(k)$ is the form factor of the potential and $k(k')$ is relative 4-momentum. 

In this paper, to solve the FY equation, the rank-one separable potential  with a Yamaguchi-type form factor is used\\
 in the monopole (Y)
\begin{eqnarray}
 g(k) = 1 / ( k^2 -  \beta^2  + i0),
\end{eqnarray}
 and dipole (Y2)
\begin{eqnarray}
 g(k) = 1 / ( k^2 -  \beta^2  + i0)^2
\end{eqnarray}
form.
    
The parameters of the potential $\lambda $ and $\beta$ were chosen so that the calculated two-particle observables coincided with the corresponding experimental data. The following observables are used: the deuteron binding energy $E_d$, the scattering length $a_0$ and  $pn$-scattering phases, and the effective radius $r_0$.

The parameters for the potential Y, as well as the results of calculating the observables, are given in the work~\cite{Rupp:1987cw}.
The parameters for the potential Y2 are given in Table 1, and the results of the calculation of the observables are presented in Table 2 and Fig. 1.

\begin{figure}[!htb]
\center{
\begin{tabular}{cc}    
	\includegraphics[width=0.45\linewidth,angle=0]{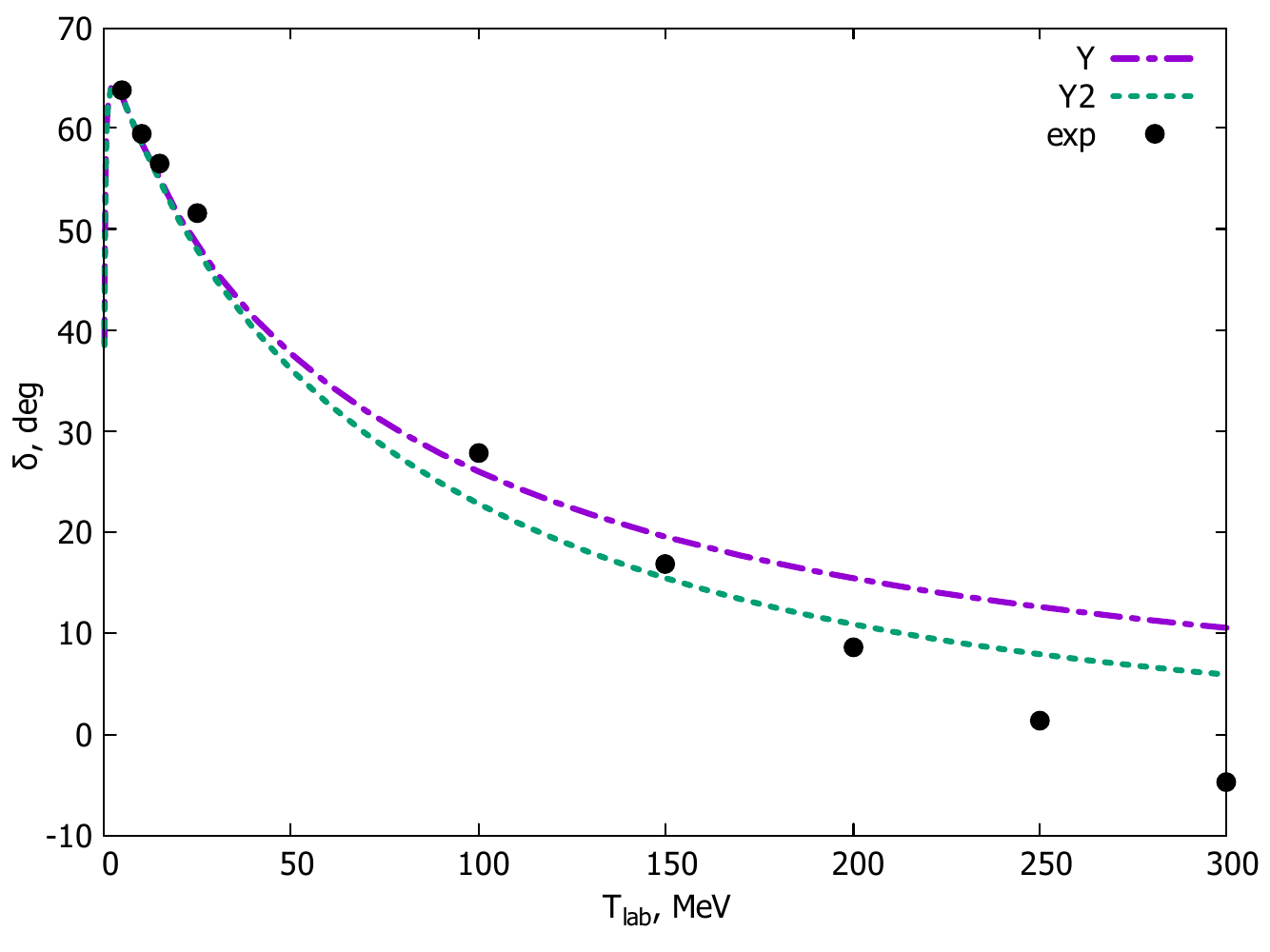}&
	\includegraphics[width=0.45\linewidth,angle=0]{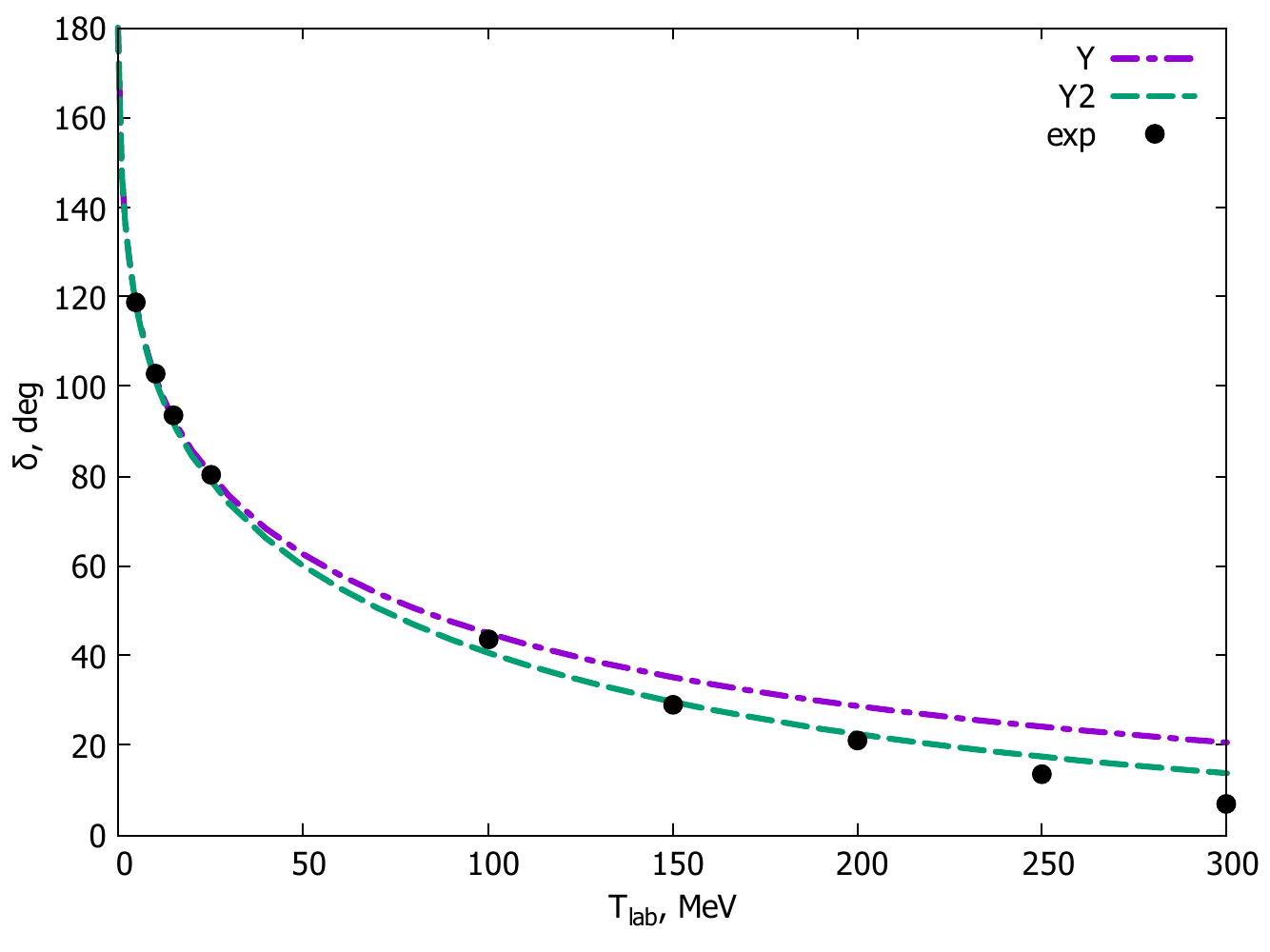}
\end{tabular}
}
\caption{Phase shifts of $pn$-scattering for $^1S_0$ (left panel) and
$^3S_1$ (right panel) calculated using the monopole and dipole Yamaguchi potentials.
}
\label{fig1}
\end{figure}

\begin{table}[ht]
\centering
\caption{Parameters of the relativistic potential $NN$-interaction for solving the BS equation}
\begin{tabular}{|l|l|c|c|}
\hline\hline
& Parameter  & Y                    &  Y2 \\ 
\hline 
$^1S_0$ & $\beta$ (GeV)        &  0.228302    &  0.336 \\
        & $\lambda$ (GeV$^4$)  & -1.12087     &  -0.071436$^a$ \\
\hline
$^3S_1$ & $\beta$ (GeV)        &  0.279731    &  0.4  \\
        & $\lambda$ (GeV$^4$)  & -3.1548      &  -0.3857451$^a$ \\
\hline\hline
\end{tabular}\\
$^a$ GeV$^8$
\end{table}

\begin{table}[ht]
\centering
\caption{Values of observables for the dipole potential of $NN$-interaction. Experimental values for the deuteron binding energy, scattering length and effective radius:  $E_d$ = 2.2246 MeV, $a_0$ = 5.424 Fm, $r_0$ = 1.756 Fm ($^3S_1$); $a_0$ = -23.748 Fm, $r_0$ = 2.75 Fm ($^1S_0$).}
\begin{tabular}{|l|c|c|}
\hline\hline 
Quantity	 &    $^1S_0$    &  $^3S_1$    \\ 
\hline 
$E_d$ (MeV)  &    --         &  2.2264   \\
$a_0$ (Fm)   &    -23.78     &  5.423   \\
$r_0$ (Fm)   &      2.71     &  1.77    \\
\hline\hline 
\end{tabular}
\end{table}

Table 3 shows the result of calculating the binding energy of a system of three nucleons using the potentials Y and Y2 obtained by solving the BSF equation. Gaussian quadrature was used to solve the equation for calculating the integrals~\cite{Bondarenko:2020baw}. Figure 2 shows the result of the convergence of the solution using the potential Y2. It depends on the number of points of the quadrature expansion in the variables $(p_4,p)$ and the value of the upper limit of integration.
It is evident that for calculation with a relative accuracy of 1 \%, it is sufficient to integrate up to 1 GeV over 64 points of the quadrature expansion by $p_4$ and up to 0.5 GeV over 15 points by $p$.
The result presented in Table 3 was obtained on the mesh $(N_{p_4},N_p)$ = (64,15).
\begin{table}[ht]
\centering
\caption{Calculation of the binding energy of a three-nucleon system (MeV). The experimental value of the binding energy of tritium is $E_t =$ 8.48 MeV}
\begin{tabular}{|l|c|c|}
\hline\hline
State	         & Y       &  Y2 \\ 
\hline 
$^3S_1$              & 25.26   &  22.99  \\
$^1S_0$, $^3S_1$    & 11.04   &  10.24  \\
\hline\hline
\end{tabular}
\end{table}

\begin{figure}[ht]
\center{
\begin{tabular}{cc}    
	\includegraphics[width=0.45\linewidth,angle=0]{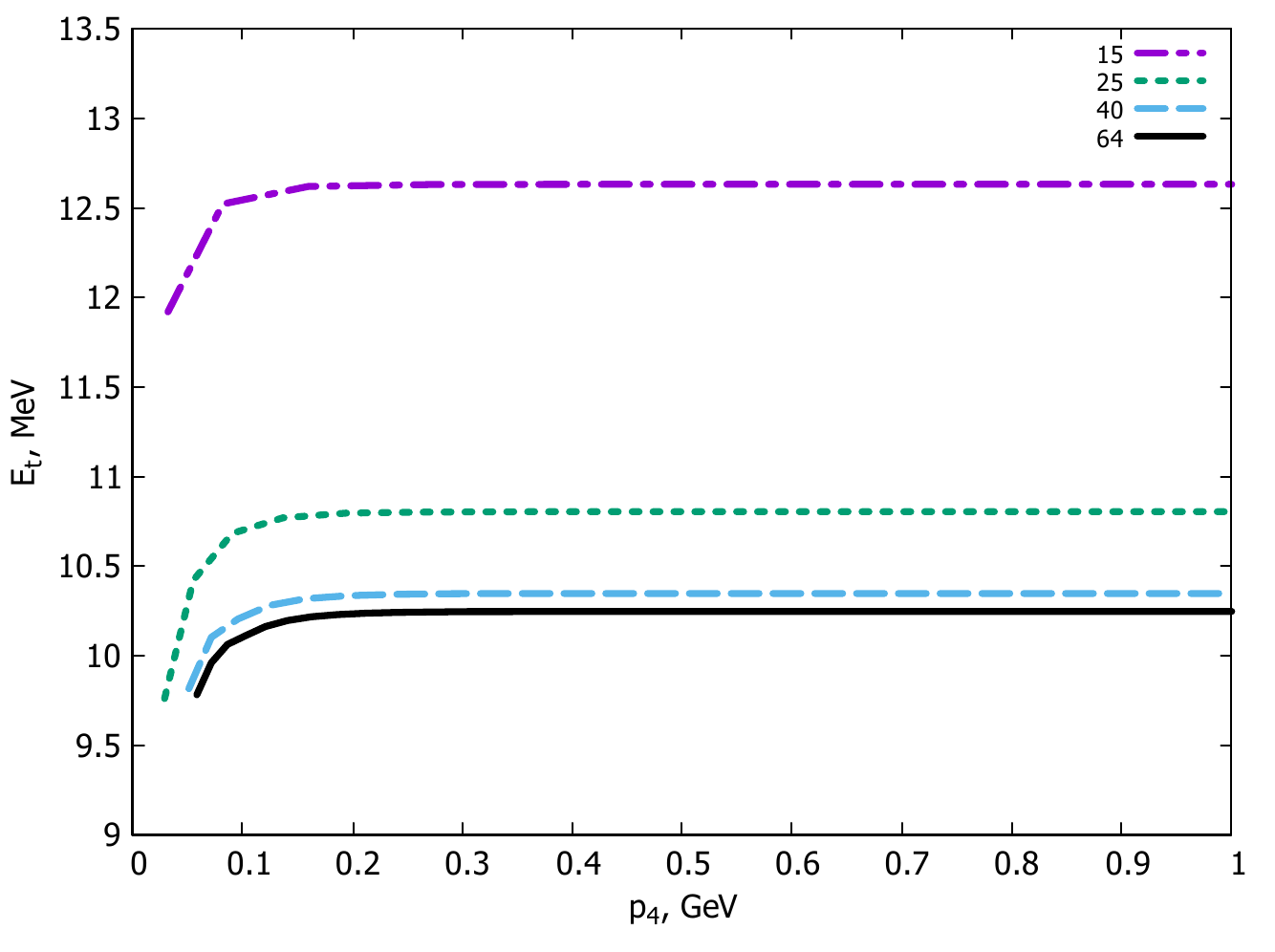}	
    	\includegraphics[width=0.45\linewidth,angle=0]{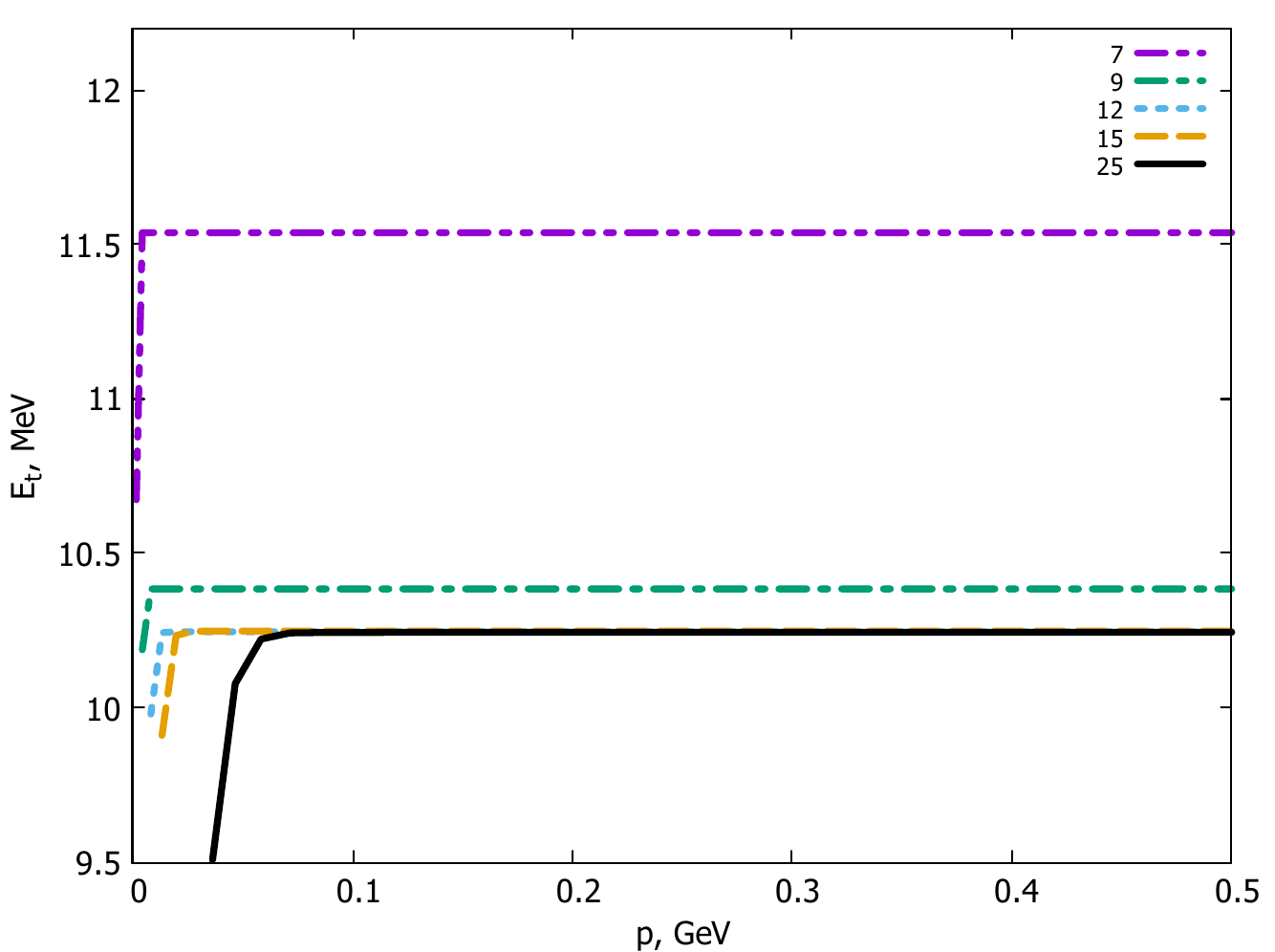}	
\end{tabular}
}
\caption{Convergence of the binding energy of a system of three particles depending on the number of points of the quadrature expansion of the integral over momentum $p_4$ (left panel) and over momentum $p$ (right panel). 
}
\label{fig2}
\end{figure}

\section{Relativistic generalization of the Faddeev-Yakubovsky equation}
The relativistic generalization of the Faddeev-Yakubovsky system of equations for a bound state of four particles in integral form in the case of a separable potential of the rank-one $NN$-interaction~(\ref{separ_NN}) was obtained 
in~\cite{Bondarenko:2024mkl} and has the following form:
\begin{eqnarray}
&&   Q_i(p,q)   = {\tau_i}(Z_{qp})  \nonumber 
 \sum_j  \int     \frac{dq'}{(2\pi)^4}    \times \\
&& [X_{ij}(p, \frac{1}{3} q + q'  )  Q_j( q  + \frac{1}{3}  q'  , q') +
 X_{ij}(p, -\frac{2}{3} q + q'  )  R_j( q  - \frac{1}{2}  q'  , q')  ] \nonumber \\  
 \nonumber \\
&&   R_i(\kappa ,s)   = 2 {\tau_i}(Z_{s}) \int     \frac{dq'}{(2\pi)^4}    
Y_{ii}(\kappa , \frac{1}{2} s + q'  )  Q_i( -s  - \frac{2}{3}  q'  , q') ,
    \label{eq_for_Q_p}
\end{eqnarray}
where   $Z_{s} = \frac{1}{2} K + s $, \ $Z_{qp} = \frac{1}{2} K + \frac{2}{3}q + p$, \
$\tau(Z)$ is the non-trivial part of the two-particle separable t-matrix, \
$K$,  $k$,  $p$,  $q$,  $\kappa$,  $s$ are the Jacobi relative momenta for a four-particle system. The functions $Q$ and $R$ refer to the schemes for constructing a system of four particles:
"3+1" \ and "2+2" \ , respectively. If the  "2+2" subchannel \ is not taken into account
then $R= 0$.

At the same time, the functions $X_{ij}$ and $Y_{ii}$ satisfy the following systems of integral equations:
	\begin{eqnarray}
&&		{X}_{ij} 
		( p,  p'  ) =  U_{ij}( p,  p'  )	+ 
  \sum_k    \int    \frac{d^4p''}{(2\pi)^4}   
		U_{ik}( p,  p''  )
		{\tau}_k(Z_{qp''}) 
		{X}_{kj}
		( p'',  p'  ) ,
   \label{eq_for_X_p}
   \\
   &&		{Y}_{ii} 		( \kappa , \kappa'  ) = 
  W_{ii} ( \kappa , \kappa' )	+ 
  \int    \frac{d^4 \kappa '' }{(2\pi)^4}   
		W_{ii} ( \kappa ,  \kappa ''  )
		{\tau_i}(Z_{s}) 
		{Y}_{ii} 
		( \kappa '',  \kappa ' ) ,
     \label{eq_for_Y_p}
	\end{eqnarray}
and the kernels $U_{ij}$ and $W_{ii}$ have the form:	
	\begin{eqnarray}
&&		U_{ij}( p,  p'  )	= C_{ij}
S(\frac{1}{4} K + \frac{1}{3} q - p')    \times          \nonumber \\
&&  S(\frac{1}{4} K + \frac{1}{3} q  + p +  p')  
		g_i(  \frac{1}{2}p + p')  	2 
		{g}_j(  p + \frac{1}{2}p') ,  \nonumber
    \\
 && 		W_{ii}( \kappa , \kappa' )	=
		S(\frac{1}{4} K + \frac{1}{2} s - \kappa')
  S(\frac{1}{4} K + \frac{1}{2} s + \kappa') 
		g_i(  \kappa)  
		{g_i}( \kappa'),  \nonumber
	\end{eqnarray}
where $C_{ij}$ is the matrix of 
recoupling coefficients  
describing the spin-isospin structure of a three-particle system~\cite{Kharchenko:1979av, Gibson:1976zz}, $S(k)$ is the scalar nucleon propagator, the indices $i$ and $j$ describe the state of pair of nucleons and 
take values $1=$ $^1S_0$ or $2=$ $^3S_1$.

\section{Numerical calculations, results and discussion}

The solution of the systems of integral equations is found by the
iteration procedure and the quadrature method is used to calculate the integrals by means of Gaussian quadrature.

First, by an iterative method on a mesh over the variables $p_0$, $q_0$, $q'_0$, $p$, $q$, $q'$, and $y = \cos \vartheta_{qq'}$, the systems of equations~(\ref{eq_for_X_p}) and~(\ref{eq_for_Y_p}) for the functions $X$ and $Y$, respectively, are solved.
Then the system of equations~(\ref{eq_for_Q_p}) is solved for the functions $Q$ and $R$, whose kernel contains the functions $X$ and $Y$ found by the solution of the systems~(\ref{eq_for_X_p}) and ~(\ref{eq_for_Y_p}).
To find function values at the points that are not mesh nodes,
for example,
$-s  - \frac{2}{3}  q'$  at each iteration, interpolation is performed.
The study of convergence was discussed in detail in our previous article~\cite{Bondarenko:2024mkl}.

The binding energy of the system of four particles $E_4$ ($K^2 = (4m_N - E_4)^2$) is found as a parameter at which there exists a solution
of the system of homogeneous integral equations~(\ref{eq_for_Q_p}) and the following condition is fulfilled:
\begin{eqnarray}
\lim_{i \to \infty}\frac{R_{i}(p,q;E_4)}{R_{i + 1}(p,q;E_4)} = 1,
\end{eqnarray}
where $i$ is the iteration number.
By varying the energy of the system it is necessary to achieve the fulfillment of this condition~\cite{Rupp:1987cw}.

Details of the calculations - transformation of the integration variables, convergence etc -
are given in~\cite{Bondarenko:2024mkl}

Since in our problem the integrals are calculated by the quadrature method,
it is obvious that the more mesh nodes are used, the more accurate the calculation result will be.
The convergence of the result of calculating the binding energy of four particles has a form similar to the three-particle case (Fig. 2)
In the previous work~\cite{Bondarenko:2024mkl} it was shown that for calculating the binding energy with an accuracy of
1 MeV, the minimum working mesh is $(N_{q_4},N_q)$ = (10,8).
In this work, the binding energy is calculated with an accuracy of 1 MeV, since the use of a separable potential of the rank-one makes more accurate calculations meaningless.

The main result of the work is the result of calculating the binding energy of the helium-4 nucleus
 illustrated by Tables 4 and 5.
 Calculations were performed on the mesh $(N_{q_4},N_q)$ = (11,9).
The experiment gives $E_4$ = 28.3 MeV for the binding energy of helium-4.
It is seen from the tables that
the results of the calculation with the potential Y and Y2 differ by
2-7 MeV depending on the state and the subamplitudes taken into account.
It is also seen that taking into account the ''2+2'' subamplitude
reduces the binding energy by 2-7 MeV in most cases.

Thus, the inclusion of the "2+2" subchannel contributes to the binding energy. And for this reason, it should be taken into account in the future  more accurate calculations of the binding energy and other observable quantities.

\begin{table}[h]
\centering
\caption{Binding energy (in MeV) of helium-4 for the Yamaguchi separable $NN$-interaction potential of monopole type}
\begin{tabular}{|l|c|c|}
	\hline
State                   & Non-relativistic calculation & Relativistic calculation \\ \hline 
$^3S_1$  w/o  ''2+2''             & 47      &   58  \\
$^1S_0$, $^3S_1$  w/o ''2+2''     & 19      &   26 \\
$^3S_1$   with   ''2+2''          & 75      &   51  \\
$^1S_0$, $^3S_1$ with   ''2+2''   & 34      &   24 \\
	\hline
\end{tabular}
\end{table}

\begin{table}[h]
\centering
\caption{Binding energy (in MeV) of helium-4 for the Yamaguchi separable $NN$-interaction potential of dipole type}
\begin{tabular}{|l|c|}
	\hline
State	     & Relativistic calculation\\ \hline 
$^3S_1$ w/o ''2+2''             &   51 \\
$^1S_0$, $^3S_1$ w/o ''2+2''    &   24 \\
$^3S_1$  with ''2+2''           &   45 \\
$^1S_0$, $^3S_1$ with   ''2+2'' &   22 \\
	\hline
\end{tabular}
\end{table}

\section{Summary}

In the paper, the relativistic generalization of the Faddeev-Yakubovsky equation has been used to calculate the binding energy of the helium-4 nucleus. An iteration method for solving the system of integral equations was used.
The rank-one separable potential of $NN$-interaction was used.
In this case, the ''2+2'' subchannel was taken into account, as a result of which the calculated binding energy changed by some MeV.

In the context of further development of the formalism under consideration, it is planned to use multirank potentials of $NN$-interaction.
The amplitudes of the state obtained by solving the equation can also be used to calculate the charge form factor of helium-4.

\section*{CRediT authorship contribution statement}  
\textbf{Serge Bondarenko:} Conceptualization, Methodology, Software, Validation, Writing – original draft, Writing – review \& editing. \textbf{Sergey Yurev:} Conceptualization, Methodology, Software, Validation, Writing – original draft, Writing – review \& editing.

\section*{Declaration of Competing Interest}

The authors declare that they have no known competing financial interests or personal relationships that could have appeared to influence the work reported in this paper.

\bibliographystyle{elsarticle-num}

\end{document}